\documentclass[12pt,preprint]{aastex}
\shorttitle{Large scale anisotropies of UHECRs}
\shortauthors{The Pierre Auger Collaboration}

\begin{document}
\slugcomment{Published in ApJ as doi: 10.1088/0004-637X/802/2/11}
\title{Large scale distribution of ultra high energy cosmic rays detected at the Pierre Auger Observatory with zenith angles up to 80$^\circ$}
\author{
{\bf The Pierre Auger Collaboration}\\
\begin{small}
A.~Aab$^{42}$, 
P.~Abreu$^{64}$, 
M.~Aglietta$^{53}$, 
E.J.~Ahn$^{81}$, 
I.~Al Samarai$^{29}$, 
I.F.M.~Albuquerque$^{17}$, 
I.~Allekotte$^{1}$, 
J.~Allen$^{84}$, 
P.~Allison$^{86}$, 
A.~Almela$^{11,\: 8}$
% ...
\\ \vspace{0.5cm}
$^{1}$ Centro At\'{o}mico Bariloche and Instituto Balseiro (CNEA-UNCuyo-CONICET), San 
Carlos de Bariloche, 
Argentina \\
% ...
$^{8}$ Instituto de Tecnolog\'{\i}as en Detecci\'{o}n y Astropart\'{\i}culas (CNEA, CONICET, UNSAM), 
Buenos Aires, 
Argentina \\
% ...
$^{11}$ Universidad Tecnol\'{o}gica Nacional - Facultad Regional Buenos Aires, Buenos Aires,
Argentina \\
% ...
$^{17}$ Universidade de S\~{a}o Paulo, Instituto de F\'{\i}sica, S\~{a}o Paulo, SP, 
Brazil \\
% ...
$^{29}$ Institut de Physique Nucl\'{e}aire d'Orsay (IPNO), Universit\'{e} Paris 11, CNRS-IN2P3, 
France \\
% ...
$^{42}$ Universit\"{a}t Siegen, Siegen, 
Germany \\
% ...
$^{53}$ Osservatorio Astrofisico di Torino  (INAF), Universit\`{a} di Torino and Sezione INFN, 
Torino, 
Italy \\
% ...
$^{64}$ Laborat\'{o}rio de Instrumenta\c{c}\~{a}o e F\'{\i}sica Experimental de Part\'{\i}culas - LIP and  
Instituto Superior T\'{e}cnico - IST, Universidade de Lisboa - UL, 
Portugal \\
% ...
$^{81}$ Fermilab, Batavia, IL, 
USA \\
% ...
$^{84}$ New York University, New York, NY, 
USA \\
% ...
$^{86}$ Ohio State University, Columbus, OH, 
USA \\
% ...
(\ddag) Deceased \\
(a) Now at Konan University \\
(b) Also at the Universidad Autonoma de Chiapas on leave of absence from Cinvestav \\
(d) Now at Unidad Profesional Interdisciplinaria de Ingenier\'{\i}a y Tecnolog\'{\i}as
Avanzadas del IPN, M\'{e}xico, D.F., M\'{e}xico \\
(e) Now at Universidad Aut\'{o}noma de Chiapas, Tuxtla Guti\'{e}rrez, Chiapas, M\'{e}xico \\
(f) Also at Vrije Universiteit Brussels, Belgium \\
\end{small}
}

\begin{abstract}
We present the results of an analysis of the large angular scale distribution of the arrival directions of cosmic rays with energy above 4 EeV detected at the Pierre Auger Observatory including for the first time events with zenith angle between $60^\circ$ and $80^\circ$. We perform two Rayleigh analyses, one in the right ascension and one in the azimuth angle distributions, that are sensitive to modulations in right ascension and  declination, respectively. The largest departure from isotropy appears in the $E > 8$ EeV energy bin, with an amplitude for the first harmonic in right ascension $r_1^\alpha =(4.4 \pm 1.0){\times}10^{-2}$, that has a chance probability $P(\ge r_1^\alpha)=6.4{\times}10^{-5}$, reinforcing the hint previously reported with vertical events alone.
\end{abstract}


\section{Introduction}

The distribution of the arrival directions of cosmic rays, together with the spectrum and composition indicators, are the main observables to try to understand their origin and nature. The dipolar component of the large scale distribution of cosmic rays has been measured by different experiments at energies below $10^{17}$ eV \citep{tibet05,tibet09,sk,milagro,eastop,ic,ic12,it,kscgde}, and has been searched for at higher energies by \citet{agasa} and the Pierre Auger Observatory. In the EeV ($\equiv 10^{18}$ eV) range the estimation of the large scale anisotropies can be useful to understand the transition from a Galactic to an extragalactic cosmic ray origin. The first hints of a change in the phase of the modulation in the right ascension distribution of arrival directions, happening around 1 EeV, are indeed suggested by the observations \citep{auger-ls2d,icrc-ls2d}. At the highest energies, the presence of a significant dipole in the extragalactic cosmic ray distribution is a likely possibility. In particular, a dipolar flux could result from cosmic rays propagating diffusively in the extragalactic turbulent magnetic fields. This could happen  if the amplitude of the field is large and/or if the cosmic rays have a component with large electric charge \citep{ha14}. A large angular scale anisotropy in the arrival direction distribution is also expected in the case that magnetic deflections are small if the cosmic ray sources are distributed similarly to the matter in the universe, due to the fact that in our local neighborhood matter is distributed inhomogeneously. These inhomogeneities lead in particular to the non-vanishing acceleration of the Local Group which is responsible for the peculiar velocity that gives rise to the observed dipole of the Cosmic Microwave Background (CMB) \citep{erdogdu06}. In fact, the non-isotropic distribution of the nearby extragalactic cosmic ray sources would lead to an excess of flux towards the direction with the highest concentration of nearby sources and this would contribute to the dipolar component of the large scale distribution of arrival directions. The maximum redshift from which extragalactic cosmic rays can arrive at Earth progressively decreases as the energy threshold increases. This is a consequence of the energy losses due to pair production and photopion production by interactions with CMB photons in the case of protons, and to photodisintegration with the CMB and infrared (IR) backgrounds in the case of heavier nuclei \citep{gr66,zk66}. Thus, the overall contribution of nearby sources becomes increasingly more important as the energy increases, leading to a larger expected anisotropy at higher energies.

The Pierre Auger Observatory has reported studies of the flux modulation in right ascension \citep{auger-ls2d,icrc-ls2d} and in both declination and right ascension \citep{auger-ls3d,auger-ls3dl,icrc-ls3d} from the analysis of events with zenith angles smaller than $60^\circ$. Upper limits on the low $\ell$ multipolar amplitudes have also been reported from a joint analysis of the Pierre Auger Observatory and the Telescope Array data, taking advantage of the full sky coverage \citep{lsTA}. In this paper we present an extension of the Pierre Auger Observatory studies including also for the first time inclined events with zenith angles between $60^\circ$ and $80^\circ$. Given the location of the Pierre Auger Observatory at a latitude $-35.2^\circ$, events arriving with zenith angles up to $60^\circ$ cover sky directions with declinations $\delta \leq 24.8^\circ$, corresponding to a fraction of 71$\%$ of the sky. By extending the zenith range up to $80^\circ$, declinations up to  $\delta \leq 44.8^\circ$ are observed, extending the accessible fraction of the sky to 85$\%$. 
 
Large angular scale modulations of the flux are studied by performing two Rayleigh analyses, one on the right ascension and another on the azimuth distribution, that are sensitive to modulations in the right ascension and declination respectively. This method is particularly useful to analyze the combined vertical plus horizontal data set as it is insensitive to small spurious modulations in the exposure as a function of the zenith angle, that could result from a difference between the energy calibration of the vertical and horizontal events. 

\section{Pierre Auger Observatory and Data Set}

The Pierre Auger Observatory \citep{nim2004} consists of an array of 1660 water-Cherenkov detectors covering 3000 km$^2$ on a triangular grid of mostly 1.5 km spacing, the surface detector (SD). It also has 4 sites with 27 telescopes overlooking the array to observe the fluorescence light emitted by the showers \citep{fd}, which allows a calorimetric measurement of the shower energy deposited in the atmosphere and is thus particularly useful for the calibration of the SD energy reconstruction. In contrast to the surface detector, the fluorescense detector (FD) has a smaller duty cycle of $13\%$.

\subsection{Data Set}
 
\begin{figure}[t]
\epsscale{.80}
\plotone{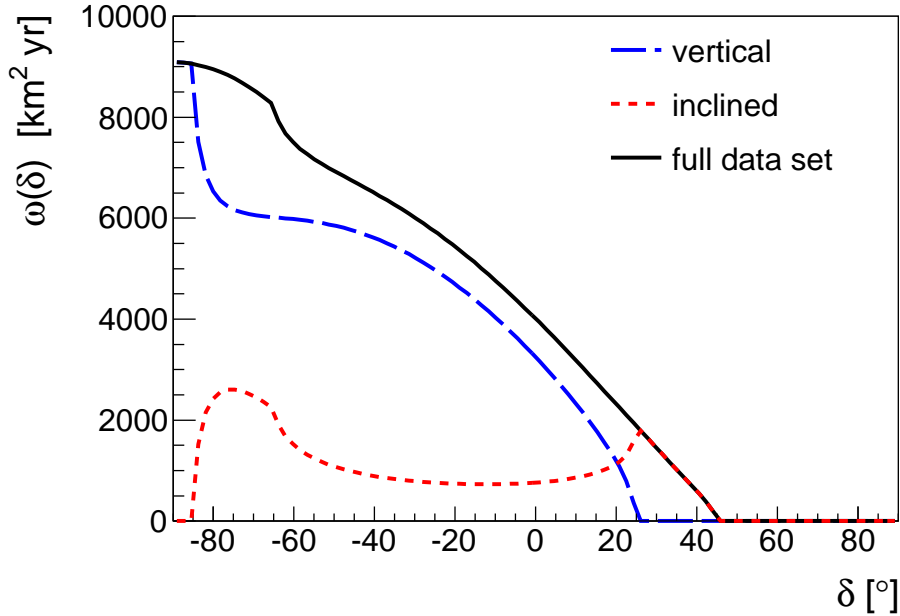}
\caption{Directional exposure as a function of the declination, computed as in  \citet{sommers2001}. The long-dashed blue line corresponds to the vertical events, the short-dashed red one to the inclined events and the solid black line to the full data set.}
\label{exposure}
\end{figure}

In this work, events recorded with the SD from 2004 January 1 to 2013 December 31 with zenith angle up to $80^\circ$ are analyzed. The quality cut imposed on events with $\theta \le 60^{\circ}$ requires that all six neighbors of the water-Cherenkov detector with the largest signal be active at the time the event was recorded. In the case of events with $\theta > 60^{\circ}$ the condition is defined differently and requires instead that the station nearest to the reconstructed core and its six neighbors be active. We also remove periods of instability on the data acquisition to have a reliable estimate of the detection exposure. The total geometric exposure, that applies to energies above full efficiency of the SD detector,  is 48,029 km$^2$ sr yr in this period. 
The directional exposure as a function of the declination is shown in Figure \ref{exposure} for events with zenith angle smaller than $60^\circ$, hereafter referred to as vertical events, for events with zenith from $60^\circ$ to $80^\circ$, referred to as inclined events, and for all events. For vertical events full efficiency is attained at 3 EeV \citep{auger-acc2010}, while for inclined events,  it is attained at 4 EeV. We will restrict the analysis to events with $E\ge 4$ EeV for which trigger effects are absent.

The event direction is determined from a fit to the arrival times of the shower front at the surface detectors. The angular resolution depends on the number of stations involved in the event. For the energies considered in this study it is always better than $0.8^\circ$. The energy reconstruction procedure is different for events above and below $60^\circ$. For vertical events the shower size at 1000 m from the shower axis, $S(1000)$, is used. From $S(1000)$ the surface energy estimator $S_{38}$, corresponding to the signal that would have been measured had the shower arrived with a zenith angle of $38^\circ$, is obtained using the constant intensity cut method \citep{auger2008}. The $S_{38}$ energy estimator is calibrated  to the energy measured by the fluorescence detector for a subset of events detected by both the surface detector and the fluorescence one. The energy resolution is better than $17\%$ \citep{icrc11-enrec}. The constant intensity cut method exploits the fact that for full efficiency and an isotropic flux the arrival direction distribution ${\rm d}N/{\rm d}\sin^2\theta$ should be constant. As discussed in Appendix A of \citet{auger-ls3d} a small deviation of this behavior, proportional to $(1+d_z\sin\ell_{\rm{obs}}\cos\theta)$, is expected when a dipolar component along the Earth rotation axis $d_z$ is present for an observation latitude $\ell_{\rm{obs}}$. This small modulation in the zenith angle distribution is not accounted for in this analysis. However, as it does not affect the distribution in azimuth nor in right ascension, which are the basis of the large scale anisotropy analysis performed, it does not affect the results presented in this paper. Inclined showers require a specific energy reconstruction method because they are dominated by muons at ground. This method is based on the fact that the shape of the muon distribution is universal for a given shower direction and that only the overall normalization of the muon distribution depends on the shower energy. This allows us to define the energy estimator $N_{19}$ as the overall normalization of a particular event with respect to a reference muon distribution, conventionally chosen to be the average muon density for primary protons of $10^{19}$ eV simulated with QGSJetII-03. Once the shower arrival direction is obtained, $N_{19}$ is reconstructed by fitting the measured signals at the surface stations to the expected muon patterns \citep{inclined}. Then, the energy of the cosmic rays is calibrated using a sub-sample of events reconstructed with both the fluorescence and surface array techniques, similarly to what is used to calibrate vertical events. The average energy resolution is $19.3\%$.  The systematic uncertainty in the energy scale associated with the fluorescence detector energy assignment, applying to both vertical and inclined events, is $14\%$ \citep{icrc-enres}.

For $E \ge 4$ EeV the number of inclined events is 15,747, while  that of vertical events is 54,467. The resulting ratio between the inclined and vertical integrated flux is $0.289 \pm 0.003$. Meanwhile, the expected ratio for a fully efficient detector and an isotropic flux is 0.293. The consistency of these ratios indicates that the energy calibrations of both data sets are compatible. This is expected as both energy estimators are calibrated with the energy measured by the fluorescence detector. 

\subsection{Atmospheric and Geomagnetic Field Effects}

As the amplitudes of the large scale modulations to be measured are rather small, at the few percent level, it is very important to carefully account for spurious effects that can modulate the flux. Variations in the array effective size due to the deployment and dead times of the detectors are taken into account by introducing a weighting factor in the Rayleigh analysis, as discussed in the next section. Furthermore, due to the steepness of the energy spectrum, even small changes in the energy estimator as a function of time or the local angular coordinates would distort significantly the counting rate of events above a given energy. In particular, the atmospheric conditions affect the shower size $S(1000)$ due to two effects. As a larger (smaller) pressure corresponds to a larger (smaller) column density traversed, an air shower will be at a more (less) advanced stage of development when it arrives at the ground. Also the air density affects the Moli\`ere radius and hence the lateral profile of the showers. These atmospheric effects are here accounted for by correcting the energy estimator of vertical events, $S(1000)$,  according to the weather conditions present at the time each event was recorded  \citep{auger-weather}. If not accounted for, the weather variations would bias the energy assignments typically by $\pm 1\%$ between the hot and cold periods of each day, and hence could affect the rates from opposite sides of the sky by up to about $\pm 2\%$ during a day, affecting the determination of the dipolar component in the direction orthogonal to the Earth rotation axis, $d_\perp$. However, once averaged over several years, strong cancellations take place and the net effect of accounting for the weather corrections is to remove a spurious $d_\perp$ component of about $0.5\%$.

The atmospheric conditions  mainly affect the electromagnetic component of the showers, that is prominent in showers with zenith angles below $60^\circ$. For the more inclined showers the muonic component is dominant and the atmospheric effects are hence expected to be negligible. We have checked this assumption by measuring the flux modulation as a function of the solar time, where no intrinsic modulation of the flux is expected but where spurious modulations due to weather conditions are maximized. No significant solar modulation is indeed observed in inclined showers and thus no weather correction is applied to showers with zenith angles above $60^\circ$.

Another effect that influences the shower size at 1000 m is the deflection of the shower particles in the geomagnetic field. Such deflections break the circular symmetry of the shower around its axis and lead to an azimuthal modulation of $S(1000)$, as has been studied in detail  for events with $\theta < 60^\circ$ in \citet{auger-geo}. If not taken into account in the energy estimator, this would induce an azimuthally dependent bias on the energy determination, leading to a spurious pseudo-dipolar pattern in the flux above a given energy threshold. This spurious dipolar component would point along the Earth's rotation axis with an amplitude $d_z$ of about $2\%$ when events with zenith angles up to $60^\circ$ are considered \citep{auger-geo}. In order to account for this effect and get an unbiased energy estimator, the measured shower size signal $S(100 0)$ is related to the one that would have been observed in the absence of the geomagnetic field, and the latter is used to construct $S_{38}$ \citep{auger-geo}. The reconstruction of events with $\theta > 60^\circ$ takes into account the geomagnetic field effect already in the expected muon distributions used to reconstruct the energy estimator $N_{19}$, and thus no further correction is needed for the inclined events.

\section{Modified Rayleigh Method}
\label{mrm}

When combining two different data sets covering different regions of the sky, such as the vertical and inclined samples considered here, a small difference in the energy cross-calibration of the samples could give rise to a difference in the measured fluxes in those regions, that could translate in a spurious large scale modulation. We will hence adopt a method that is essentially insensitive to these effects, studying the large scale distribution of the arrival directions by performing a classical Rayleigh analysis \citep{linsley75} over both the right ascension and the azimuth angle distributions. The analysis is slightly generalized by weighting each event by a factor that takes into account small modulations in the exposure arising from the variations in the operating size of the array as a function of time, and for the effects of a small net tilt of the array surface \citep{auger-ls3d}. 

The number of active detector cells $n_{\rm{cell}}(t)$ (number of active detectors having their six neighbors active) is constantly monitored at the Observatory. The total number of active cells, $N_{\rm{cell}}$, as a function of the sidereal time $\alpha_0$ (measured by the right ascension of the zenith at the center of the array) and its relative variations, $\Delta N_{\rm{cell}}$, are obtained from
\begin{equation}
N_{\rm{cell}} (\alpha_0)=\sum_j n_{\rm{cell}}(\alpha_0+j\ T_{\rm{sid}}), \hspace{1cm} \Delta N_{\rm{cell}} (\alpha_0)=\frac{N_{\rm{cell}} (\alpha_0)}{\langle N_{\rm{cell}} \rangle},
\end{equation}
with $\langle N_{\rm{cell}} \rangle = T_{\rm{sid}}^{-1} \int_0^{T_{\rm{sid}}} {\rm d}\alpha_0 N_{\rm{cell}}(\alpha_0)$, where $T_{\rm{sid}}$ corresponds to the duration of the sidereal day. The small modulations in right ascension of the flux induced by these variations is accounted for by weighting each event by a factor $w_i \propto \Delta N_{\rm{cell}}^{-1}(\alpha_0^i)$. The modulation in the total period of time considered has an amplitude of $0.24\%$, with the phase of the maximum at $\alpha_0 = 44^\circ$. If not accounted for this modulation would lead to a spurious dipole component $d_\perp \sim 0.2\%$. Note that the corresponding modulation at the solar frequency has instead a much larger amplitude of $3.5\%$, and it is the cancellation along the years, for 10 years of continuous operation of the Observatory, that leads to the small resulting amplitude at the sidereal frequency.

The geometric aperture of a horizontal array is given by $N_{\rm{cell}} (\alpha_0)\ a_{\rm{cell}}(\theta)$, where $a_{\rm{cell}} (\theta) = 1.95 \cos\theta\ {\rm km}^2$ \citep{auger-acc2010}. However, the fact that the height above sea level of the array of detectors has a small average tilt of about $0.2^\circ$ towards a direction $30^\circ$ from the East to the South ($\phi_{\rm{tilt}}=-30^\circ$) modulates the effective cell area according to
\begin{equation}
a_{\rm{cell}}(\theta,\phi)=1.95 [1+0.003 \tan \theta \cos (\phi-\phi_{\rm{tilt}})] \cos \theta.
\end{equation}
For energies above full efficiency the tilt effect can be taken into account by including in the weight of each event a factor $[1+0.003 \tan \theta \cos (\phi-\phi_{\rm{tilt}})]^{-1}$ and neglecting the modulation in $\phi$ in the exposure. If not accounted for the tilt would lead to a spurious dipole component $d_z \sim 0.2\%$.

The Fourier coefficients of the modified Rayleigh analysis in right ascension  ($\alpha_i$ of each event) are then given by 
\begin{equation}
\label{abkra}
a_k^\alpha=\frac{2}{\mathcal{N}}\sum_{i=1}^N w_i\cos(k \alpha_i), 
\hspace{1cm}
b_k^\alpha=\frac{2}{\mathcal{N}}\sum_{i=1}^N w_i\sin(k \alpha_i),
\end{equation}
where the sums run over the number of events $N$ in the considered energy range, the weights are given by $w_i\equiv [\Delta N_{\rm{cell}}(\alpha^i_0) (1+0.003 \tan \theta_i \cos (\phi_i-\phi_{\rm{tilt}}))]^{-1}$, and the normalization factor is $\mathcal{N}=\sum_{i=1}^Nw_i$. The amplitude $r_k^\alpha$ and phase $\varphi_k^\alpha$ of the event rate modulation are estimated as
\begin{equation}
\label{eqn:fh2}
r_k^\alpha=\sqrt{(a_k^\alpha)^2+(b_k^\alpha)^2}, \hspace{1cm}\varphi_k^\alpha=\frac{1}{k}\arctan\frac{b_k^\alpha}{a_k^\alpha}.
\end{equation}
The weight factors $w_i$ are very close to 1 in the present analysis, and thus the probability $P(\ge r_k^\alpha)$ that an amplitude equal to or larger than $r_k^\alpha$ arises from an isotropic distribution can be safely approximated by the cumulative distribution function of the Rayleigh distribution $P(\ge r_k^\alpha)=\exp{(-\kappa_0)}$, where $\kappa_0=\mathcal{N}(r_k^\alpha)^2/4$. 

The Fourier coefficients for the Rayleigh analysis in azimuth are given by the same expressions, just changing $\alpha$ by $\phi$. Notice that after having accounted for the modulation induced by the tilt and the geomagnetic effect, the azimuthal distribution is expected to be uniform for energies above full efficiency for an isotropic distribution of cosmic rays. The amplitude $b_1^\phi$ is actually sensitive to asymmetries between the northern and southern local flux, and thus gives information on the dipolar component along the Earth's rotation axis.

We restrict the analysis to the first two harmonics $k=1,2$. The first harmonic coefficients in right ascension and azimuth are enough to reconstruct the dipole in the hypothesis that the higher order multipole contributions are negligible, as will be done in Section \ref{dipolerec}. The second harmonic coefficients ($k=2$) are sensitive to the quadrupole component (and higher order multipoles) of the cosmic ray distribution. The presence of an equatorial dipole component leads to non-vanishing Rayleigh coefficients $a_1^\alpha$ and/or $b_1^\alpha$ and hence to a non-vanishing amplitude $r_1^\alpha$. In general, in an expansion in spherical harmonics ($\Phi({\delta,\alpha})=\sum_{\ell,m} a_{\ell m}Y^{\ell m}({\pi/2-\delta,\alpha})$), all the terms $a_{\ell m}$ with $m=\pm k$ contribute to the $a_k^\alpha$ and $b_k^\alpha$ coefficients. Then, when neglecting $a_{\ell m}$ with $\ell >1$, the two Rayleigh coefficients $a_1^\alpha$ and $b_1^\alpha$ are sufficient to determine the two multipoles $a_{1 \pm1}$. However if we want to also reconstruct the quadrupole, neglecting only the $a_{\ell m}$ with $\ell >2$, then the four Rayleigh coefficients  $a_1^\alpha$, $b_1^\alpha$, $a_2^\alpha$ and $b_2^\alpha$ are not sufficient to determine the six multipoles  $a_{1 \pm1}$, $a_{2 \pm1}$ and $a_{2 \pm2}$. The missing information can be recovered by considering also the first order Rayleigh coefficients of the events coming from  the southern hemisphere and from the northern hemisphere separately, as discussed in the Appendix. Finally the $a_{\ell 0}$ coefficients can be obtained from the Rayleigh analysis in azimuth up to order $\ell$. 

We consider energies above the full efficiency of inclined events, splitting them in two bins, 4 to 8 EeV and $E > 8$ EeV, updating the results for the large scale anisotropy for the two highest energy bins reported in \citet{auger-ls2d,auger-ls3d,icrc-ls2d,icrc-ls3d}  with a larger sky coverage and nearly twice the number of events.

\subsection{Right Ascension Distribution}

In this section we present the results for the Rayleigh coefficients in right ascension and we will discuss the determination of the dipole in the next section. In particular, $a_1^\alpha$ and $b_1^\alpha$ will be used to reconstruct the equatorial dipole in Section \ref{dipolerec}, while $a_2^\alpha$ and $b_2^\alpha$ probe the quadrupole.

The results for the modified Rayleigh analysis are quoted in Table \ref{tab:raharmonics} including the $a_k^\alpha$ and $b_k^\alpha$ coefficients with their statistical uncertainty $\sigma=\sqrt{2/{\mathcal{N}}}$, the amplitude $r_k^\alpha$ and phase $\varphi_k^\alpha$, as well as the probability that a larger or equal amplitude arises by chance from an isotropic distribution.

\begin{table}[ht]
\begin{center}
\caption{Rayleigh analysis in right ascension}
\begin{tabular}{ c c c c c c c c }
\tableline\tableline
$E$ [EeV] & $N$ & $k$ & $a_k^\alpha$ & $b_k^\alpha$ & $r_k^\alpha$ & $\varphi_k^\alpha$ &
$P(\ge r_k^\alpha)$ \\
\tableline
4 - 8  & 50,417 & 1 &$\phantom{-}0.0030 \pm 0.0063$ & $\phantom{-}0.0008 \pm 0.0063$ & 0.0031 & $15^\circ$ & 0.88 \\ 
  &  &2 & $-0.0012 \pm 0.0063$ & $-0.0004 \pm 0.0063$ & 0.0013 & $99^\circ$ & 0.98 \\ 
\\
$> 8$ & 19,797 & 1 &$-0.004 \pm 0.010$ & $0.044 \pm 0.010$ & 0.044 & $95^\circ$ & $6.4{\times}10^{-5}$\\ 
 &  &2 & $\phantom{-}0.009 \pm 0.010$ & $0.027 \pm 0.010$ & 0.028 & $36^\circ$ & 0.021\\ 
\tableline
\end{tabular}
\label{tab:raharmonics}
\end{center}
\end{table}

\begin{figure}[t]
\epsscale{1.80}
\centerline{\includegraphics[angle=0,width=10cm]{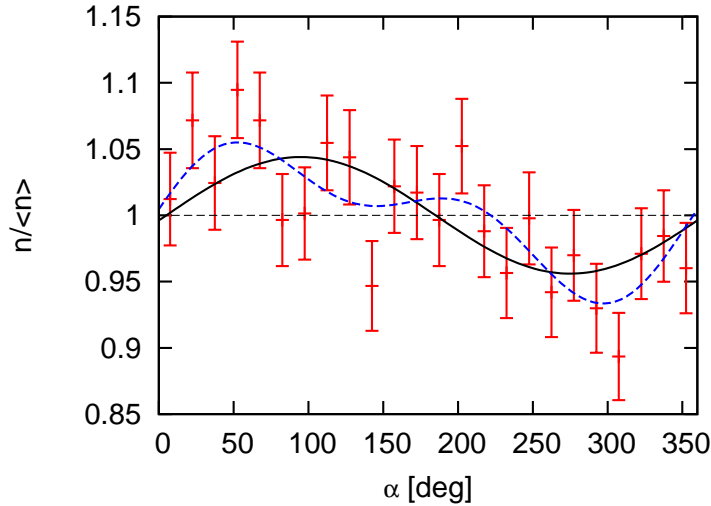}}
\caption{Observed number of events over the mean as a function of the right ascension with 1 $\sigma$ error bars for $E >$ 8 EeV. The black solid line shows the first harmonic modulation from Table \ref{tab:raharmonics}, while the blue dashed line shows the combination of the first and second harmonics. }
\label{nra}
\end{figure}

In the lower energy bin, between 4 and 8 EeV, all the coefficients are consistent with zero within their uncertainties, and there is no evidence for departures from isotropy in the right ascension distribution. In the higher energy bin, $E > 8$ EeV, the first harmonic has an amplitude $r_1^\alpha = 0.044 \pm 0.010$, with a chance probability to arise from an isotropic distribution of $P(\ge r_1^\alpha) = 6.4{\times}10^{-5}$. The phase $\varphi_1^\alpha$ points to $95^\circ$. Both the amplitude and the phase are in agreement with previous measurements reported in \citet{auger-ls2d,icrc-ls2d}. Due to the larger statistics, arising both from the larger time period considered as well as from the inclusion of the inclined events with $60^\circ < \theta < 80^\circ$, the significance of the measurement has grown to about $4 \sigma$. The amplitude of the second harmonic is less significant, with a $2\%$ probability to arise by chance. We show in Figure \ref{nra} the ratio of the observed number of events to the mean number as a function of the right ascension, together with the first harmonic and the first plus second harmonics results.

A useful test to check if the systematic effects are well controlled is to repeat the analysis at the solar and the antisidereal frequencies. Each sidereal day is slightly shorter than the solar day by about 4 minutes,  so that the sidereal year has 366.25 days. The antisidereal time is an artificial time scale in which the day is longer than a solar day by about 4 minutes, and therefore has 364.25 days per year. The weather and array size variations have the largest effect in producing spurious modulations at the solar frequency where the effects are not cancelled under the integration over several full years. No physical phenomena are expected to occur in the antisidereal frequency, however the combination of solar and seasonal systematic distortions could produce a spurious modulation in the antisidereal time. We report in Table \ref{solarantisid} the  amplitude of the Fourier transform of the arrival times of the events  obtained after applying the weather correction and weighting the events with the factor to account for the modulation of the number of active detectors at the solar  (365.25 cycles/year) and antisidereal  (364.25 cycles/year) frequencies. No signs of spurious effects appear for any of the  energy bins.

\begin{table}[ht]
\begin{center}
\caption{First harmonic analysis in solar and antisidereal frequencies}
\begin{tabular}{ c c c c c c }
\tableline\tableline
& $E$ [EeV] & $r_1$ & $\varphi_1$ [h] & $P(\ge r_1)$ \\
\tableline
solar & 4 - 8  & $0.0110 \pm 0.0063$ & 14 &  0.21 \\ 
      & $> 8$ & $0.005 \pm 0.010$ & 17 & 0.86 \\ 
\\
antisidereal & 4 - 8 & $0.0046 \pm 0.0063$ & 8 & 0.76 \\
  & $> 8$ & $0.017 \pm 0.010$ & 13 & 0.24\\
\tableline
\end{tabular}
\label{solarantisid}
\end{center}
\end{table}

As a check that no large weather effect is present in the inclined events data set ($\theta > 60^\circ$), we also performed the Rayleigh analysis at the solar frequency for all inclined events with  $E\ge~4$ EeV. The  amplitude obtained is $r_1^{\rm{solar}}= 0.012 \pm 0.011$, showing no sign of the presence of a  weather modulation.

\subsection{Azimuth Distribution}

A dipolar component of the flux along the rotation axis of the Earth gives rise to a non-vanishing $b_1^\phi$ coefficient. Moreover, in general, each $b_k^\phi$ coefficient with odd $k$ and each $a_k^\phi$ coefficient with even $k$ receives contributions from all of the $a_{\ell 0}$ multipole coefficients with $\ell \ge k$ in a spherical harmonics expansion ($\Phi({\delta,\alpha})=\sum_{\ell,m} a_{\ell m}Y^{\ell m}({\pi/2-\delta,\alpha})$).  On the other hand, the $a_k^\phi$ coefficients with odd $k$ and the $b_k^\phi$ with even $k$ probe asymmetries between the eastern and western directions, that are expected to be zero when many full sidereal days are integrated. The results of the Rayleigh analysis in the azimuth angle are reported in Table \ref{phiharmonics}.

\begin{table}[ht]
\begin{center}
\caption{Rayleigh analysis in azimuth}
\begin{tabular}{ c c c c c c l l }
\hline
$E$ [EeV] & $N$ & $k$ & $a_k^\phi$ & $b_k^\phi$ & $P(\ge |a_k^\phi|)$ &  $P(\ge |b_k^\phi|)$ \\
\tableline\tableline
4 - 8  & 50,417 & 1&$-0.0116 \pm 0.0063$ & $-0.0142 \pm 0.0063$ & 0.064 & 0.024\\
  &  &2& $-0.0034 \pm 0.0063$ & $-0.0066 \pm 0.0063$ & 0.59 & 0.29 \\ 
\\
$> 8$ & 19,797 & 1&$-0.009 \pm 0.010$ & $-0.024 \pm 0.010$ & 0.35 & 0.015\\
 &  &2& $-0.006 \pm 0.010$ & $\phantom{-}0.008 \pm 0.010$ & 0.58 & 0.45 \\ 
\tableline
\end{tabular}
\label{phiharmonics}
\end{center}
\end{table}
The largest departure from isotropy appears for the $b_1^\phi$ coefficient in both energy bins, although with low statistical significance ($2.4\%$ and $1.5\%$ probability, respectively).  The  $a_2^\phi$ coefficient that probes the quadrupolar component is subdominant (and compatible with zero) in both energy bins. The  $a_1^\phi$ and $b_2^\phi$ coefficients are compatible with zero, as expected. 

\section{Dipole Reconstruction}

In this section the reconstruction of the dipole components from the Rayleigh coefficients obtained in the last section is performed, first in the simplified approximation that only the dipole contribution to large scale anisotropies is relevant, which is justified by the fact that the $k=2$ coefficients determined in the previous section are not significantly different from zero. Then the reconstruction is performed  considering also a possible quadrupole contribution. The reconstruction of the dipole (and quadrupole) components through this method does not require a precise knowledge of the directional acceptance of vertical and inclined events, that would depend on the relative energy calibration of both samples. A miscalibration of one of the samples would just lead to a slight shift of the energy bins to which the events contribute, but without introducing a spurious modulation in right ascension or azimuth that could affect the determination of the dipole components.

\subsection{Dipolar Pattern}
\label{dipolerec}

A pure dipolar anisotropy can be parametrized as a function of the arrival direction ${\hat u}$ as
\begin{equation}
\label{phidip}
\Phi({\hat u})=\frac{\Phi_0}{4\pi}(1+{\vec d}\cdot {\hat u}).
\end{equation}
The observed arrival direction distribution is obtained by convoluting the flux with the detector exposure $\omega({\hat u})$, giving
\begin{equation}
\label{dNdO}
\frac{{\rm d}N}{{\rm d}\Omega}({\hat u})=\Phi({\hat u})\,\omega({\hat u}).
\end{equation}
As a function of the local coordinates ($\theta$,$\phi$,$\alpha_0$) the exposure $\omega$ can be considered to be a function of $\theta$ only, as the effects of the small modulation in $\phi$ and $\alpha_0$ are already accounted for in the modified Rayleigh analysis. Assuming a general dipole with maximum amplitude $d$ in the right ascension and declination direction ($\alpha_d$,$\delta_d$), and writing the angular dependence of the flux in terms of local coordinates\footnote{Using the fact that ${\hat d}\cdot{\hat u}=\sin\delta_d(\cos\theta \sin\ell_{\rm{obs}}+\sin\theta \cos\ell_{\rm{obs}} \sin\phi)+\cos\delta_d\cos\alpha_d(-\sin\theta \cos\phi \sin\alpha_0+\cos\theta \cos\ell_{\rm{obs}} \cos\alpha_0-\sin\theta \sin\ell_{\rm{obs}} \sin\phi \cos\alpha_0)+\cos\delta_d \sin\alpha_d(\sin\theta \cos\phi \cos\alpha_0+\cos\theta \cos\ell_{\rm{obs}} \sin\alpha_0-\sin\theta \sin\ell_{\rm{obs}} \sin\phi \sin\alpha_0).$}, the first harmonic amplitudes in $\phi$ can be expressed by means of  integrals of the flux as
\begin{eqnarray}
a_1^\phi&=&\frac{2}{\mathcal{N}}\int_0^{2\pi}{\rm d}\alpha_0\int_0^{2\pi}{\rm d}\phi \int_{\theta_{\rm{min}}}^{\theta_{\rm{max}}}{\rm d}\theta \sin\theta \cos\phi\, \Phi(\theta,\phi,\alpha_0)=0,\label{a1phid}\\
b_1^\phi&=&\frac{2}{\mathcal{N}}\int_0^{2\pi}{\rm d}\alpha_0\int_0^{2\pi}{\rm d}\phi \int_{\theta_{\rm{min}}}^{\theta_{\rm{max}}}{\rm d}\theta \sin\theta \sin\phi\, \Phi(\theta,\phi,\alpha_0)=\frac{\pi}{\mathcal{N}}\Phi_0 d_z \cos\ell_{\rm{obs}} {\overline {\sin\theta}},\label{b1phid}\\
{\mathcal{N}}&=&\int_0^{2\pi}{\rm d}\alpha_0\int_0^{2\pi}{\rm d}\phi \int_{\theta_{\rm{min}}}^{\theta_{\rm{max}}}{\rm d}\theta \sin\theta\, \Phi(\theta,\phi,\alpha_0)=\pi \Phi_0 ({\overline 1}+d_z \sin\ell_{\rm{obs}} {\overline {\cos\theta}}),
\end{eqnarray}
where in the last terms the integrals over $\phi$ and $\alpha_0$ have been performed, $d_z$ is the dipole component along the Earth's rotation axis, $d_z=d\sin\delta_d$, $\ell_{\rm{obs}}$ is the latitude of the Observatory, and we denoted by ${\overline {f(\theta)}}\equiv\int_{\theta_{\rm{min}}}^{\theta_{\rm{max}}}{\rm d}\theta \sin\theta f(\theta)$.
The coefficient $a_1^\phi$ vanishes as anticipated, while $b_1^\phi$ is related to $d_z$ by
\begin{equation}
b_1^\phi=\frac{d_z \cos\ell_{\rm{obs}} \langle \sin\theta\rangle}{1+d_z \sin\ell_{\rm{obs}} \langle \cos\theta\rangle},
\end{equation}
where we have used that ${\overline {\sin\theta}}/{\overline 1}$ can be estimated as the mean value of $\sin(\theta)$ of the events themselves, $\langle \sin\theta\rangle$, and similarly ${\overline {\cos\theta}}/{\overline 1} \simeq \langle \cos\theta\rangle$. Finally, for $d_z \sin\ell_{\rm{obs}} \langle \cos\theta\rangle \ll 1$, the dipole component along the Earth's rotation axis can be obtained to linear order as $d_z=b_1^\phi/(\cos\ell_{\rm{obs}} \langle \sin\theta\rangle)$.

On the other hand, the equatorial component of the dipole can be recovered from the Rayleigh analysis in right ascension, to linear order in the dipole amplitude, through $d_\perp \simeq r_1^\alpha/\langle\cos\delta\rangle$, where $\langle\cos\delta\rangle$ is the mean cosine declination of the events \citep{auger-ls2d}. 

The resulting dipole components from the Rayleigh coefficients determined in the last section are reported in Table \ref{dipolecomp}. The dipole component along the Earth's rotation axis $d_z$, the equatorial component $d_\perp$, the total amplitude $d$, as well as the direction $(\alpha_{\rm{d}},\delta_{\rm{d}})$  are quoted for the two energy bins.
\vspace{3mm}
\begin{table}[ht]
\begin{center}
\caption{Dipole components and directions in equatorial coordinates.}
\begin{tabular}{ c c c c c c c c l }
\tableline\tableline
$E$ [EeV] & $d_z$ & $d_\perp$ & $d$ &  $\delta_{\rm{d}}$ & $\alpha_{\rm{d}}$    \\
\tableline
4 - 8  & $-0.027 \pm 0.012$  & $0.004\pm 0.008$ & $0.027 \pm 0.012$&$-81^\circ \pm 17^\circ$&$15^\circ\pm 115^\circ $ \\ 
$> 8$&$-0.046\pm 0.019$&$0.057\pm 0.013$&$0.073\pm 0.015$&$-39^\circ\pm 13^\circ$&$95^\circ\pm 13^\circ$\\ 
\tableline
\end{tabular}
\label{dipolecomp}
\end{center}
\end{table}

All of the dipole components obtained in both energy bins are compatible with the ones previously reported in \citet{auger-ls3d,icrc-ls3d} within the systematic uncertainties. The dipole amplitude in the higher energy bin is also consistent with the upper limit to the dipole amplitude at $99\%$ CL reported by the joint analysis of the Auger and TA data at energies above 8.5 EeV \citep{lsTA}. These bounds depend on the dipole direction in the sky and range from $8\%$ for directions close to the equator to $13\%$ for directions close to the poles.  

\subsection{Dipole and Quadrupole Patterns}

Assuming now that the angular distribution of the flux can be well approximated by the combination of a dipole plus a quadrupole, it can be parametrized as
\begin{equation}
\label{phidipquad}
\Phi({\hat u})=\frac{\Phi_0}{4\pi}\left(1+{\vec d}\cdot {\hat u}+\frac{1}{2}\sum_{i,j}Q_{ij}u_iu_j\right),
\end{equation}
with $Q_{ij}$ the symmetric and traceless quadrupole tensor. From the measured values of $b_1^\phi$ and $a_2^\phi$ obtained from the Rayleigh analysis in $\phi$ performed in the previous section, $d_z$ and $Q_{zz}$ can be determined through Eqs.~(\ref{dz}) and (\ref{qzz}), as discussed in the Appendix. From the right ascension Rayleigh coefficients $a_2^\alpha$ and $b_2^\alpha$ (and taking into account that $Q_{ij}$ is traceless) the quadrupole coefficients $Q_{xy}$, $Q_{xx}$ and $Q_{yy}$ can be determined through Eqs.~(\ref{qxy}) and (\ref{qxxyy}) in the Appendix.

As $a_1^\alpha$ results from a combination of contributions from $d_x$ and $Q_{xz}$, and $b_1^\alpha$ from a combination of $d_y$ and $Q_{yz}$, two more independent measurements are needed to determine the four parameters. As discussed in the Appendix, a simple way of separating $d_x$ and $Q_{xz}$ is through computing $a_1^\alpha$ for the southern and northern subsamples of events, $a_1^{\alpha S}$ and $a_1^{\alpha N}$, obtained by restricting the sums in Eq.~(\ref{abkra}) to events with $\delta <0$ and $\delta >0$, respectively. Similarly, $d_y$ and $Q_{yz}$ can be separated by measuring  $b_1^{\alpha S}$ and $b_1^{\alpha N}$.

In Table \ref{raharmonicsns} we report the first harmonics in right ascension for the events coming from the southern and northern hemispheres for the two energy bins considered.
\begin{table}[ht]
\begin{center}
\caption{First harmonic in right ascension for events arriving from the southern and northern hemispheres.}
\begin{tabular}{ c c c c c c c c }
\tableline\tableline
$E$ [EeV] & Hem & $N$ & $a_1^\alpha$ & $b_1^\alpha$ & $r_1^\alpha$ & $\varphi_1^\alpha$ &
$P(\ge r_1^\alpha)$ \\
\tableline
4 - 8  & S & 40,256 & $0.0034 \pm 0.0070$ & $-0.0010 \pm 0.0070$ & 0.0036 & $344^\circ$ & 0.88 \\ 
  & N &10,161  & $0.001 \pm 0.014$ & $0.008 \pm 0.014$ & 0.008 & $79^\circ$ & 0.85 \\ 
\\
$> 8$ & S & 15,878 &$-0.005 \pm 0.011$ & $0.042 \pm 0.011$ & 0.042 & $96^\circ$ & $7.9{\times}10^{-4}$\\ 
 & N & 3919 & $-0.001 \pm 0.022$ & $0.051 \pm 0.022$ & 0.051 & $91^\circ$ & 0.075\\ 
\tableline
\end{tabular}
\label{raharmonicsns}
\end{center}
\end{table}

In the energy bin between 4 and 8 EeV the amplitude in both hemispheres is compatible with zero within the uncertainties. This means that the fact that the $r_1^\alpha$ amplitude for the full data set vanishes as reported in Table \ref{tab:raharmonics} is not due to a cancellation of two significant and opposite modulations in the northern and the southern hemispheres. For $E>8$ EeV the modulation is more significant and has the same phase in both hemispheres, indicating that the dipolar contribution to the modulation dominates over the quadrupolar one. 

\begin{table}[ht]
\begin{center}
\caption{Reconstruction with dipole and quadrupole patterns}
\begin{tabular}{ c l l }
\tableline\tableline
$E$ [EeV] & $d_i$ &  $Q_{ij}$   \\
\tableline
4 - 8  & $ d_z=-0.012\pm 0.030$  & $Q_{zz}=0.028 \pm 0.052$   \\ 
&$d_x=0.003 \pm 0.010$  &$Q_{xx}=-0.018 \pm 0.032$ \\
&$d_y=0.005 \pm 0.010$   & $Q_{xy}=-0.001 \pm 0.019$ \\
&  &$Q_{xz}=-0.004 \pm 0.024$  \\
&  & $Q_{yz}=0.013 \pm 0.024$ \\
\\
$> 8$&$d_z=-0.021\pm 0.048$& $Q_{zz}=0.046 \pm 0.083$ \\ 
 &$d_x=-0.003 \pm 0.016$ &$Q_{xx}=0.004 \pm 0.051$ \\
 &$d_y=0.055 \pm 0.016$ &$Q_{xy}=0.080 \pm 0.030$ \\
 & &$Q_{xz}=0.007 \pm 0.039$ \\
 & &$Q_{yz}=-0.004 \pm 0.039$ \\
\tableline
\end{tabular}
\label{dipolequadcomp}
\end{center}
\end{table}
Table \ref{dipolequadcomp} reports the dipolar and quadrupolar reconstructed components. In both energy bins the reconstructed dipolar components are consistent with those reported in Table \ref{dipolecomp} in the hypothesis of a pure dipolar anisotropy.  Note that in Table \ref{dipolecomp} $d_\perp$ is consistent with 0 in the energy bin from 4 to 8 EeV, and so are $d_x$ and $d_y$ in Table \ref{dipolequadcomp}. For E$>8$ EeV, $\alpha_d$ is very close to $90^\circ$ in Table \ref{dipolecomp}, and so $d_x\simeq 0$ and $d_y\simeq d_\perp$.
The most significant quadrupole component is the $Q_{xy}$ one in the $E>8$ EeV bin, that according to Eq.~(\ref{qxy}) is proportional to the second harmonic in right ascension $b_2^\alpha$, whose amplitude has a 2$\%$ probability to arise by chance from isotropy (see Table~\ref{tab:raharmonics}).

\begin{figure}[ht]
\epsscale{.80}
\plotone{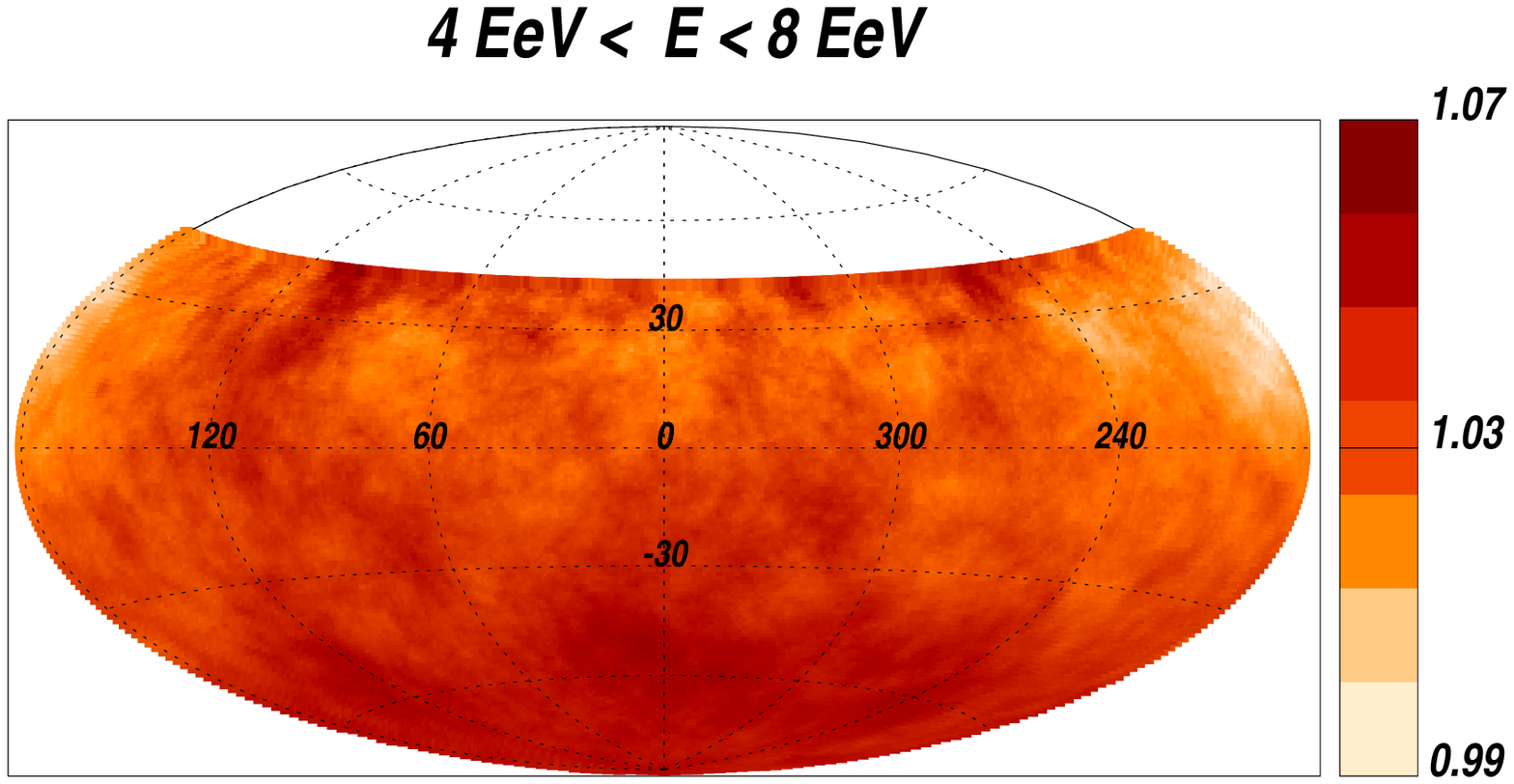}
\plotone{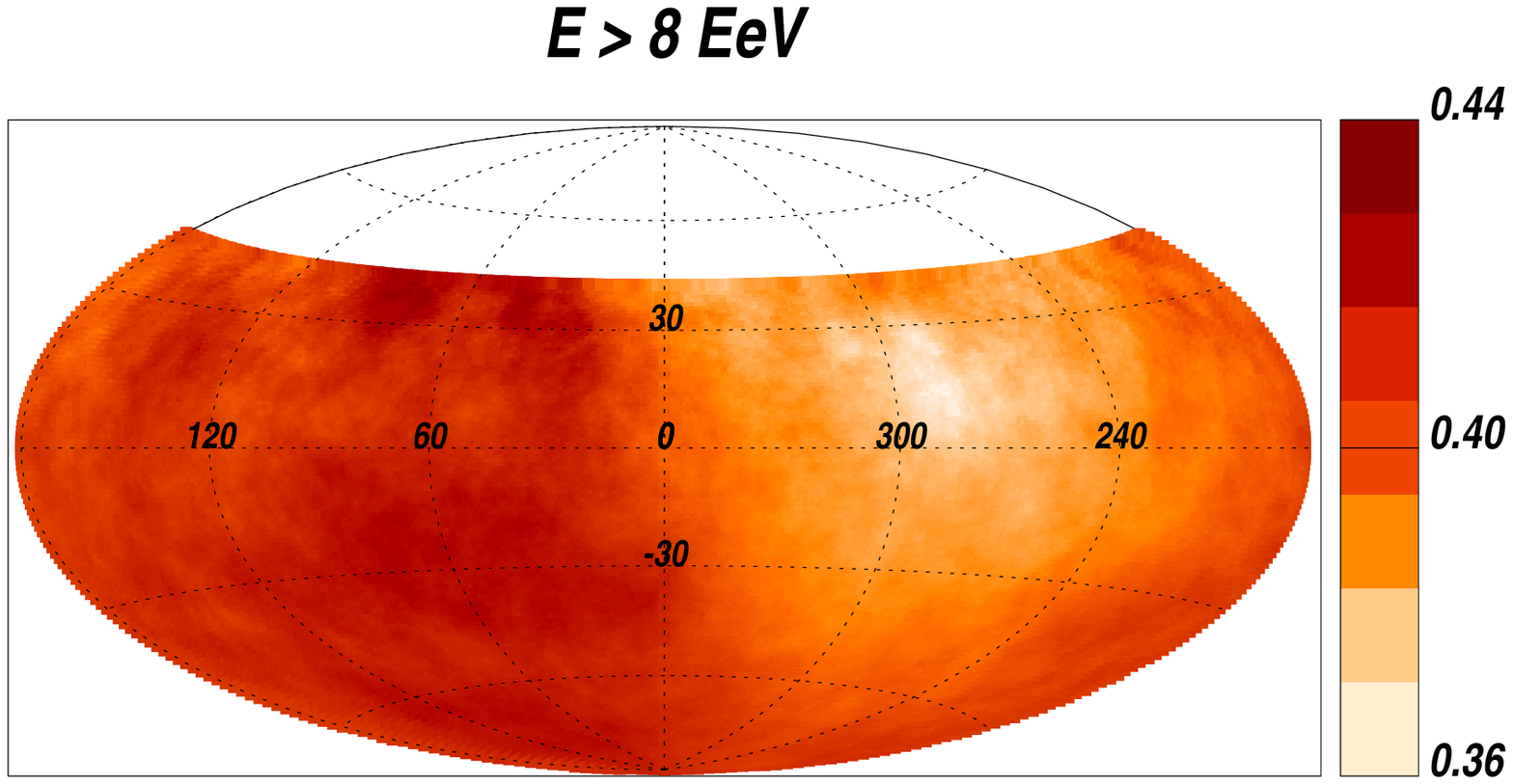}
\caption{Sky map in equatorial coordinates of flux, in km$^{-2}$ yr$^{-1}$ sr$^{-1}$ units, smoothed in angular windows of $45^\circ$ and for the two energy bins.  }
\label{map}
\end{figure}

We show in Fig.~\ref{map} the sky maps in equatorial coordinates of the flux of cosmic rays, in units of km$^{-2}$ yr$^{-1}$ sr$^{-1}$, smoothed in an angular window of $45^\circ$ for the two energy bins considered. The upper panel corresponds to the energy bin between 4 and 8 EeV, while the lower panel corresponds to $E > 8$ EeV. Notice the difference in the color scales of flux variations appearing in the two plots. While for the high energy bin the maximum flux is 21$\%$ larger than the minimum one, for the lower energy bin this ratio is just 8$\%$.

\section{Conclusions}

We presented the results of an analysis of the large angular scale distribution of the arrival directions of the Pierre Auger Observatory data including for the first time inclined events with zenith angle between $60^\circ$ and $80^\circ$. The inclusion of the inclined events not only provides an increase of about 30$\%$ in the number of events, but also leads to a larger fraction of the sky covered, up to 85$\%$. We performed two Rayleigh analyses, in the right ascension and azimuth angles, that are sensitive to the right ascension and declination modulation of the flux, respectively. Two energy bins above the full efficiency for inclined events were analyzed: from 4 to 8 EeV and above 8 EeV. No significant departure from isotropy is observed in the distribution of events in the energy bin between 4 and 8 EeV. For energies above 8 EeV the first harmonic in right ascension has an amplitude  $r_1^\alpha =(4.4 \pm 1.0){\times}10^{-2}$ with a chance probability $P(\ge r_1^\alpha)=6.4{\times}10^{-5}$, reinforcing the hint reported in \citet{icrc-ls2d} with vertical events alone detected up to the end of 2012. 

The Rayleigh analysis in azimuth, sensitive to modulations in the declination direction, gives first harmonic coefficients $b_1^\phi=-0.014\pm 0.006$ for energies between 4 and 8 EeV and  $b_1^\phi=-0.024\pm 0.010$ for energies larger than 8 EeV. The negative values in both energy bins correspond to a dipolar component $d_z$ pointing to the South, although the amplitudes have low statistical significance, with chance probabilities of $2.4\%$ and $1.5\%$, respectively. 

Under the assumption that the only significant contribution to the anisotropy is from the dipolar component, the observations above 8 EeV correspond to a dipole  of amplitude $d = 0.073\pm 0.015$ pointing to $(\alpha,\delta)= (95^\circ\pm 13^\circ,-39^\circ\pm 13^\circ)$. If a quadrupolar contribution is also included, the resulting dipole is consistent with that obtained in the previous case, although with a larger uncertainty, and the quadrupole components obtained are not significant.

\acknowledgments

The successful installation, commissioning, and operation of the Pierre Auger Observatory would not have been possible without the strong commitment and effort from the technical and administrative staff in Malarg\"{u}e. 

We are very grateful to the following agencies and organizations for financial support: 
Comisi\'{o}n Nacional de Energ\'{\i}a At\'{o}mica, Fundaci\'{o}n Antorchas, Gobierno de la Provincia de Mendoza, Municipalidad de Malarg\"{u}e, NDM Holdings and Valle Las Le\~{n}as, in gratitude for their continuing cooperation over land access, Argentina; the Australian Research Council; Conselho Nacional de Desenvolvimento Cient\'{\i}fico e Tecnol\'{o}gico (CNPq), Financiadora de Estudos e Projetos (FINEP), Funda\c{c}\~{a}o de Amparo \`{a} Pesquisa do Estado de Rio de Janeiro (FAPERJ), S\~{a}o Paulo Research Foundation (FAPESP) Grants No.2012/51015-5, 2010/07359-6 and No. 1999/05404-3, Minist\'{e}rio de Ci\^{e}ncia e Tecnologia (MCT), Brazil; Grant No. MSMT-CR LG13007, No. 7AMB14AR005, No. CZ.1.05/2.1.00/03.0058 and the Czech Science Foundation Grant No. 14-17501S, Czech Republic;  Centre de Calcul IN2P3/CNRS, Centre National de la Recherche Scientifique (CNRS), Conseil R\'{e}gional Ile-de-France, D\'{e}partement Physique Nucl\'{e}aire et Corpusculaire (PNC-IN2P3/CNRS), D\'{e}partement Sciences de l'Univers (SDU-INSU/CNRS), Institut Lagrange de Paris (ILP) Grant No. LABEX ANR-10-LABX-63, within the Investissements d'Avenir Programme  Grant No. ANR-11-IDEX-0004-02, France; Bundesministerium f\"{u}r Bildung und Forschung (BMBF), Deutsche Forschungsgemeinschaft (DFG), Finanzministerium Baden-W\"{u}rttemberg, Helmholtz Alliance for Astroparticle Physics (HAP), Helmholtz-Gemeinschaft Deutscher For\-schungs\-zen\-trum (HGF), Helmholtz Alliance for Astroparticle Physics (HAP), Ministerium f\"{u}r Wissenschaft und Forschung, Nordrhein Westfalen, Ministerium f\"{u}r Wissenschaft, For\-schung und Kunst, Baden-W\"{u}rttemberg, Germany; Istituto Nazionale di Fisica Nucleare (INFN), Ministero dell'Istruzione, dell'Universit\`{a} e della Ricerca (MIUR), Gran Sasso Center for Astroparticle Physics (CFA), CETEMPS Center of Excellence, Italy; Consejo Nacional de Ciencia y Tecnolog\'{\i}a (CONACYT), Mexico; Ministerie van Onderwijs, Cultuur en Wetenschap, Nederlandse Organisatie voor Wetenschappelijk Onderzoek (NWO), Stichting voor Fundamenteel Onderzoek der Materie (FOM), Netherlands; National Centre for Research and Development, Grants No. ERA-NET-ASPERA/01/11 and No. ERA-NET-ASPERA/02/11, National Science Centre, Grants No. 2013/08/M/ST9/00322, No. 2013/08\-/M/ST9/00728 and No. HARMONIA 5 - 2013/10/M/ST9/00062, Poland; Portuguese national funds and FEDER funds within Programa Operacional Factores de Competitividade through Funda\c{c}\~{a}o para a Ci\^{e}ncia e a Tecnologia (COMPETE), Portugal; Romanian Authority for Scientific Research ANCS, CNDI-UEFISCDI partnership projects Grants No. 20/2012 and No. 194/2012, Grants No. 1/ASPERA2/2012 ERA-NET, No. PN-II-RU-PD-2011-3-0145-17 and No. PN-II-RU-PD-2011-3-0062, the Minister of National  Education, Programme  Space Technology and Advanced Research (STAR), Grant No. 83/2013, Romania; Slovenian Research Agency, Slovenia; Comunidad de Madrid, FEDER funds, Ministerio de Educaci\'{o}n y Ciencia, Xunta de Galicia, European Community 7th Framework Program, Grant No. FP7-PEOPLE-2012-IEF-328826, Spain; Science and Technology Facilities Council, United Kingdom; Department of Energy, Contracts No. DE-AC02-07CH11359, No. DE-FR02-04ER41300, No. DE-FG02-99ER41107 and No. DE-SC0011689, National Science Foundation, Grant No. 0450696, The Grainger Foundation, USA; NAFOSTED, Vietnam; Marie Curie-IRSES/EPLANET, European Particle Physics Latin American Network, European Union 7th Framework Program, Grant No. PIRSES-2009-GA-246806; and UNESCO.

\appendix

\section{APPENDIX: RECONSTRUCTION OF DIPOLAR AND QUADRUPOLAR COMPONENTS}

We present here the reconstruction of the dipolar and quadrupolar components in the case where the angular distribution of the flux at Earth can be well approximated by the combination of a dipole plus a quadrupole. In this case the flux can be parametrized as in eq. (\ref{phidipquad}).

 Analogously to eq. (\ref{b1phid}) in this case $b_1^\phi$ and  $a_2^\phi$ can be written by direct integration in terms of $d_z$ and $Q_{zz}$ as
\begin{eqnarray}
b_1^\phi&=&\frac{\pi}{\mathcal{N}}\Phi_0 \cos\ell_{\rm{obs}}\left(d_z {\overline {\sin\theta}} +\frac{3}{2} Q_{zz}\sin\ell_{\rm{obs}}{\overline {\sin\theta \cos\theta}}\right),\\
a_2^\phi&=&-\frac{3\pi}{8\mathcal{N}}\Phi_0 \cos^2\ell_{\rm{obs}} {\overline {\sin^2\theta}} Q_{zz}.
\end{eqnarray}
Then, from the measured values of $b_1^\phi$ and $a_2^\phi$, and using that to leading order $\mathcal{N} \simeq \pi \Phi_0 {\overline {1}}$, $d_z$ and $Q_{zz}$ can be determined as
\begin{eqnarray}
d_z&=&\frac{1}{\langle\sin\theta\rangle\cos\ell_{\rm{obs}}}\left(b_1^\phi+4 a_2^\phi\tan\ell_{\rm{obs}}\frac{\langle\sin\theta\cos\theta\rangle}{\langle\sin^2\theta\rangle}\right),\label{dz}\\
Q_{zz}&=&-\frac{8}{3}\frac{a_2^\phi}{\cos^2\ell_{\rm{obs}}\langle\sin^2\theta\rangle}.
\label{qzz}
\end{eqnarray}
The right ascension Rayleigh coefficients are also obtained from direct integration as
\begin{equation}
a_k^\alpha=\frac{2}{\mathcal{N}}\int_{\delta_{\rm{min}}}^{\delta_{\rm{max}}}{\rm d}\delta\cos\delta\, \omega(\delta)\int_0^{2\pi}{\rm d}\alpha \cos(k \alpha)\, \Phi(\delta,\alpha),
\end{equation}
where $\delta_{\rm{min}}$ and $\delta_{\rm{max}}$ are the minimum and maximum declination considered ($-90^\circ$ and $44.8^\circ$ respectively, when the full data set is considered). The coefficient $b_k^\alpha$ is given by  a similar expression changing $\cos(k \alpha)$ to $\sin(k \alpha)$. Then,
\begin{eqnarray}
a_1^\alpha&=&\frac{\Phi_0}{2\mathcal{N}}\left(d_x{\widetilde {\cos\delta}}+Q_{xz} {\widetilde {\cos\delta \sin\delta}} \right),\label{a1ra}\\
b_1^\alpha&=&\frac{\Phi_0}{2\mathcal{N}}\left(d_y{\widetilde {\cos\delta}}+Q_{yz} {\widetilde {\cos\delta \sin\delta}} \right),\label{b1ra}\\
a_2^\alpha&=&\frac{\Phi_0}{8\mathcal{N}}\left(Q_{xx}-Q_{yy}\right){\widetilde {\cos^2\delta}},\label{a2ra}\\
b_2^\alpha&=&\frac{\Phi_0}{4\mathcal{N}}Q_{xy}{\widetilde {\cos^2\delta}},\label{b2ra}
\end{eqnarray}
where we denoted ${\widetilde {f(\delta)}}\equiv \int_{\delta_{\rm{min}}}^{\delta_{\rm{max}}}{\rm d}\delta\cos\delta\, \omega(\delta) f(\delta)$, and to leading order $\mathcal{N} \simeq \Phi_0 {\widetilde {1}}/2$. From the last two equations, we obtain that
\begin{eqnarray}
Q_{xy}&=&\frac{2b_2^\alpha}{\langle \cos^2\delta \rangle},\label{qxy}\\
Q_{xx}-Q_{yy}&=&\frac{4a_2^\alpha}{\langle \cos^2\delta \rangle},\label{qxxyy}
\end{eqnarray}
where we have used that ${\widetilde {\cos^2\delta}}/{\widetilde {1}}$ can be estimated by the mean value  $\langle\cos^2\delta\rangle$ of the events. Taking into account that the quadrupole tensor is traceless, from Eqs.~(\ref{qzz}) and  (\ref{qxxyy}) the three diagonal terms can be obtained.

The $d_x$ and $Q_{xz}$ components appear combined in  $a_1^\alpha$ (and similarly $d_y$ and $Q_{yz}$ in $b_1^\alpha$), and cannot be disentangled by just measuring the first harmonic amplitudes in right ascension for the full data set, as both coefficients represent a modulation proportional to $\cos\alpha$. The difference is that the modulation induced by $d_x$ is symmetric with respect to the equatorial plane (same sign in the northern and southern hemispheres) while that induced by $Q_{xz}$ is antisymmetric (opposite  sign in the northern and southern hemispheres). Then a simple way of separating $d_x$ and $Q_{xz}$ is computing $a_1^\alpha$ for the southern and northern subsamples of events, $a_1^{\alpha S}$ and $a_1^{\alpha N}$, restricting the sums in Eq.~(\ref{abkra}) to events with $\delta <0$ and $\delta >0$, respectively. Similarly $d_y$ and $Q_{yz}$ can be separated by measuring  $b_1^{\alpha\rm{S}}$ and $b_1^{\alpha\rm{N}}$. From Eqs.~(\ref{a1ra}) and (\ref{b1ra}) we can write
\begin{eqnarray}
a_1^{\alpha\rm{S(N)}}&=& d_x  \langle \cos\delta \rangle_{\rm{S(N)}} + Q_{xz} \langle \cos\delta \sin\delta \rangle_{\rm{S(N)}},\\
b_1^{\alpha\rm{S(N)}}&=& d_y  \langle \cos\delta \rangle_{\rm{S(N)}} + Q_{yz} \langle \cos\delta \sin\delta \rangle_{\rm{S(N)}},
\end{eqnarray}
where $\langle\cdot\rangle_{\rm{S}}$ and $\langle\cdot\rangle_{\rm{N}}$ denote the mean values over the events from the southern and northern hemispheres, respectively. We can then estimate the corresponding dipolar and quadrupolar components as
\begin{eqnarray}
d_x&=&\frac{a_1^{\alpha\rm{S}} \langle \cos\delta \sin\delta \rangle_{\rm{N}} -a_1^{\alpha\rm{N}} \langle \cos\delta \sin\delta \rangle_{\rm{S}}}{\langle \cos\delta \rangle_{\rm{S}}\langle \cos\delta \sin\delta \rangle_{\rm{N}} -\langle \cos\delta \rangle_{\rm{N}}\langle \cos\delta \sin\delta \rangle_{\rm{S}}},\\
Q_{xz}&=&\frac{a_1^{\alpha\rm{S}}\langle \cos\delta \rangle_{\rm{N}}-a_1^{\alpha\rm{N}}\langle \cos\delta \rangle_{\rm{S}}}{\langle \cos\delta \rangle_{\rm{N}}\langle \cos\delta \sin\delta \rangle_{\rm{S}}-\langle \cos\delta \rangle_{\rm{S}}\langle \cos\delta \sin\delta \rangle_{\rm{N}}},
\end{eqnarray}
and 
\begin{eqnarray}
d_y&=&\frac{b_1^{\alpha\rm{S}} \langle \cos\delta \sin\delta \rangle_{\rm{N}} -b_1^{\alpha\rm{N}} \langle \cos\delta \sin\delta \rangle_{\rm{S}}}{\langle \cos\delta \rangle_{\rm{S}}\langle \cos\delta \sin\delta \rangle_{\rm{N}} -\langle \cos\delta \rangle_{\rm{N}}\langle \cos\delta \sin\delta \rangle_{\rm{S}}},\\
Q_{yz}&=&\frac{b_1^{\alpha\rm{S}}\langle \cos\delta \rangle_{\rm{N}}-b_1^{\alpha\rm{N}}\langle \cos\delta \rangle_{\rm{S}}}{\langle \cos\delta \rangle_{\rm{N}}\langle \cos\delta \sin\delta \rangle_{\rm{S}}-\langle \cos\delta \rangle_{\rm{S}}\langle \cos\delta \sin\delta \rangle_{\rm{N}}}.
\end{eqnarray}

\end{document}